\documentclass{moriond}

\usepackage{color}
\usepackage{amssymb}
\usepackage[utf8]{inputenc}

\newcommand{\be}{\begin{equation}}
\newcommand{\ee}{\end{equation}}
\newcommand{\bea}{\begin{eqnarray}}
\newcommand{\eea}{\end{eqnarray}}

\newcommand{\MKK}{M_{\rm KK}}

\def\WSS{Witten-Sakai-Sugimoto}

\definecolor{Red}{rgb}{0.7,0,0}
\definecolor{Green}{rgb}{0,0.4,0}
\definecolor{Blue}{rgb}{0,0,0.7}
\definecolor{gray}{rgb}{0.5,0.5,0.5}

\def\red{\color{red}}

\def\gray{\color{gray}}

\def\be{\begin{equation}}
\def\ee{\end{equation}}
\def\bea{\begin{eqnarray}}
\def\eea{\end{eqnarray}}

\newcommand{\Photo}{}

\begin{document}
\vspace*{4cm}
\title{SCALAR AND TENSOR GLUEBALL DECAY RATES\\
FROM THE WITTEN-SAKAI-SUGIMOTO MODEL
}

\author{ A. REBHAN }

\address{Institute for Theoretical Physics, TU Wien,\\
Wiedner Hauptstraße 8-10, A-1040 Vienna, Austria}

\maketitle\abstracts{
The holographic Witten-Sakai-Sugimoto model is an almost parameter-free
approximation to large-$N_c$ QCD with chiral quarks that in a number of cases
gives surprisingly good quantitative results for light meson physics.
It also allows for predictions of the decay rates of glueballs into pseudoscalar and
vector mesons through the coupling of
modes on the flavor branes to the massive gravitational modes in the bulk.
Including effects from finite quark masses leads to an enhancement of
the decay of scalar glueballs into heavier pseudoscalars in good agreement with the decay pattern
observed for the
scalar glueball candidate $f_0(1710)$. Tensor glueballs are found to have a large width
when their masses are above the 2$\rho$ and 2$K^*$ thresholds. With a
mass as suggested by lattice gauge theory the lowest tensor glueball appears to be too broad to be observable, 
in contrast to the pseudoscalar glueball which may be a rather narrow state.
}


Glueballs, color-neutral bound states of gluons only, are a cornerstone prediction of QCD.
They appear as additional states in the meson spectrum usually made of quark-antiquark pairs, but
because of possible mixing of flavorless $q\bar q$ states and glueballs, the experimental
status of the latter is still very unclear.\cite{Crede:2008vw}
In quenched lattice QCD, the glueball spectrum is fairly well established,\cite{Morningstar:1999rf}
with a rather mild dependence on the number of colors~\cite{Lucini:2012gg}. Moreover,
a recent study~\cite{Gregory:2012hu} points to only comparatively small changes
when dynamical quarks are included. This seems to suggest that glueballs in real QCD may be approached
from the large-$N_c$ limit, where glueballs are parametrically narrow and mixing is suppressed.

Gauge/gravity duality offers a possible avenue towards QCD in the large-$N_c$ limit. In particular, 
the Witten-Sakai-Sugimoto model~\cite{Witten:1998zw,Sakai:2004cn,Sakai:2005yt} 
is an almost parameter-free string-theoretic construction that
is connected to the low-energy limit of large-$N_c$ QCD with massless quarks, albeit the limitations
of the supergravity approximation make it prohibitively difficult to systematically improve this model.
Nevertheless, the Witten-Sakai-Sugimoto model has not only been found to successfully reproduce
several qualitative features of low-energy QCD, but frequently gives quantitative predictions within
10-30\% of experimental results.\cite{Rebhan:2014rxa}

In this model, the confinement scale is related to a Kaluza-Klein scale $\MKK$ where a five-dimensional
supersymmetric Yang-Mills theory is dimensionally reduced to non-supersymmetric pure Yang-Mills theory.
Flavor D8 branes which accommodate chiral quarks in the fundamental representation are asymptotically 
separated in the extra spatial dimension, but they connect in the bulk, thereby realizing chiral symmetry
breaking. Vector mesons appear as flavor gauge field modes on the D8 branes, relating $\MKK$ to the mass of the $\rho$ meson,
while the 't Hooft coupling at this scale is related to the pion decay constant. This gives~\cite{Sakai:2004cn,Sakai:2005yt} $\MKK=949$ MeV
and $\lambda\approx 16.63$. Matching alternatively~\cite{Rebhan:2014rxa} 
to the large-$N_c$ lattice result of the ratio of string tension to $m_\rho^2$
gives a lower value of $\lambda\approx 12.55$.
Encouragingly, with this range of parameters, the experimental decay rate of $\rho$ and $\omega$ mesons is
covered by the result from the \WSS\ model, which is of order $\lambda^{-1}N_c^{-1}$ and $\lambda^{-4}N_c^{-2}$, respectively.\cite{Sakai:2005yt}
The model also provides a Witten-Veneziano mass for the would-be U(1)$_\mathrm{A}$ Goldstone boson,\cite{Sakai:2004cn} which
combined with explicit quark masses yields masses for $\eta$ and $\eta'$ mesons that deviate only by $\sim 10\%$
from experiment.\cite{Brunner:2015oga}

The spectrum of glueballs in this model has been worked out already
2000 by Brower, Mathur, and Tan\,\cite{Brower:2000rp}. The lightest (scalar) state turns out to
correspond to an ``exotic'' graviton polarization\,\cite{Constable:1999gb}, followed
by a (mostly) dilatonic scalar and tensor mode which are degenerate in mass.
This appears to resemble the spectrum observed in lattice QCD, where the lightest glueball is
significantly lighter than the tensor glueball. However, with $\MKK$ fixed by the $\rho$ meson
of the Sakai-Sugimoto construction, the lightest (``exotic'') scalar glueball comes out at a mere
$M_E\approx 855$ MeV, whereas the mostly dilatonic scalar (and the tensor) mode has $M_D=M_T=1487$ MeV, which
is close to lattice results for the scalar glueball, but too light for the tensor.

Decay rates for the holographic glueballs have been first considered by Hashimoto, Tan, and Terashima\,\cite{Hashimoto:2007ze}.
Brünner, Parganlija, and myself~\cite{Brunner:2015oqa} have revisited these calculations and extended them beyond the lowest
(exotic) mode. Surprisingly enough, the latter turns out to be much broader than the heavier dilatonic scalar,
which adds to the suspicion\,\cite{Constable:1999gb} that the former might have to be discarded from the model.\footnote{At
best, the exotic scalar glueball could play a role as a broad glueball component of the $\sigma$ meson
as proposed by Narison\,\cite{Narison:1996fm} (albeit the $\sigma$ meson itself is absent in the Witten-Sakai-Sugimoto model).} 
The predominantly dilatonic scalar is indeed more similar to how a scalar glueball is usually represented
in gauge/gravity duality. The result for its decay rate into (massless) pions reads
\be\label{GDpipi}
\Gamma(G_D\to\pi\pi)/M_D\approx{1.359}/{\lambda N_c^2}\approx0.009\ldots0.012.
\ee
This is smaller than the experimental value $0.025(3)$ for the glueball candidates $f_0(1500)$ whose mass would
match almost perfectly with $M_D$. Moreover, this isoscalar $0^{++}$ meson decays predominantly into four pions,
whereas decay into four pions is very strongly suppressed in the holographic model.
Indeed, phenomenological models that favor $f_0(1500)$ as a glueball candidate\,\cite{Amsler:1995td}
do so with rather large mixing with $q\bar q$ states, contrary to the working hypothesis of the holographic approach.

In fact, recently attention has turned increasingly to the meson $f_0(1710)$
as a glueball candidate.\cite{Lee:1999kv,Janowski:2014ppa,Cheng:2015iaa}
Its (not officially established) decay rate into pions indeed seems to match the result (\ref{GDpipi}).
The main difficulty in accepting $f_0(1710)$ as glueball candidate is however the strong flavor asymmetry in its decay
pattern where decays into kaons dominate. To explain this, ``chiral suppression'' of scalar glueball decay
has been proposed\,\cite{Chanowitz:2005du}, but the underlying perturbative reasoning appears questionable\,\cite{Frere:2015xxa}.
In the holographic setup, explicit quark mass terms almost unavoidably lead to additional vertices between
glueballs and pseudoscalar mesons. Brünner and myself~\cite{Brunner:2015yha,Brunner:2015oga} have
studied this question in the Witten-Sakai-Sugimoto model when it is deformed by explicit mass terms
and we have found that when the explicit mass terms couple in the same manner as the Witten-Veneziano mass term
the resulting decay pattern indeed reproduces that of $f_0(1710)$ surprisingly well, see Table \ref{tabrates}.
Here we have extrapolated the mass of the glueball to the observed mass of $f_0(1710)$
in a manner which keeps the dimensionless chiral result (\ref{GDpipi}) fixed. However, because the mass of $f_0(1710)$
is above the threshold of $2\rho$ and $2\omega$ mesons, the decay into four and six pions is no longer suppressed.
Together with limits on the rate of decays into $\eta\eta'$ pairs~\cite{Brunner:2015oga}, this represents
a falsifiable prediction of the holographic model.

\def\PDG{\,\cite{PDG15}}
\def\BPR{\,\cite{Brunner:2015oqa}}
\def\BR{\,\cite{Brunner:2015yha}}
\def\BReep{\,\cite{Brunner:2015oga}}
\def\BRc{\,\cite{Brunner:2015yha,Brunner:2015oga}}

\begin{table}[p]
\caption[]{Experimental data from the Particle Data Group\PDG\ for the decay rates of the glueball candidates $f_0(1500)$ and
$f_0(1710)$ [in gray color: the latter combined
with the branching ratio\,\cite{Albaladejo:2008qa} ${\rm Br}(f_0(1710)\to KK)=0.36(12)$]
compared to the results obtained for a pure glueball $G_D$ of same mass in the chiral Witten-Sakai-Sugimoto (WSS) model\BPR\ 
and in a deformed model with finite quark masses\BR. 
Red color indicates a significant discrepancy with a pure-glueball interpretation of the respective $f_0$ meson. \newline
}
\label{tabrates}
\begin{center}
\begin{tabular}{|l|r|r|r|}
\hline
decay &  $\Gamma/M$ {(exp.\protect\PDG)}  & (WSS chiral\,\protect\cite{Brunner:2015oqa}) & (WSS massive\protect\BR)\\
\hline
$f_0(1500)$ (total) & 0.072(5)  & \red 0.027\ldots0.039
& 0.057\ldots0.079
\\
$f_0(1500)\to4\pi$ & 0.036(3)  &  \red 0.003\ldots 0.005  &  
\red 0.003\ldots 0.005  \\
$f_0(1500)\to2\pi$ & 0.025(2)  & \red 0.009\ldots0.012 & \red 0.010\ldots0.014\\
$f_0(1500)\to 2K$ & 0.006(1)  & \red 0.012\ldots0.016 & \red 0.034\ldots0.045\\
$f_0(1500)\to 2\eta$ & 0.004(1)  & 0.003\ldots0.004 & \red 0.010\ldots0.013\\
$f_0(1500)\to \eta\eta'$ & 0.0014(6)  & \red 0 & $(^{*})$ 0\\
\hline
$f_0(1710)$ (total) & 0.081(5) 
& 0.059\ldots0.076 & 0.083\ldots0.106 \\
$f_0(1710)\to 2K$ & 
\gray 0.029(10) 
& \red 0.012\ldots0.016 & 0.029\ldots0.038 \\[8pt]
$f_0(1710)\to 2\eta$ & \gray 0.014(6) 
& \red 0.003\ldots0.004 & 0.009\ldots0.011 \\[8pt]
$f_0(1710)\to2\pi$ & \gray 0.012($+5\atop-6$) 
& 0.009\ldots0.012 & 0.010\ldots0.013 \\[8pt]
$f_0(1710)\to2\rho,\rho\pi\pi\to4\pi$ & ? & 0.024\ldots 0.030 & 0.024\ldots 0.030 \\ 
$f_0(1710)\to2\omega$ & \gray 0.010($+6\atop-7$) 
& 0.011\ldots 0.014 & 0.011\ldots 0.014 \\ 
$f_0(1710)\to \eta\eta'$ & ?  & 0 & $(^{*})$ 0\\
\hline
\end{tabular}
\end{center}
\footnotesize $(^{*})$ If one relaxes\BReep\ the assumption\BR\ of a universal coupling of all pseudoscalar mass terms, nonzero $\eta\eta'$ rates are possible in the WSS model with finite quark masses. In the case of $f_0(1710)$ an upper limit of $\Gamma(\eta\eta')/\Gamma(\pi\pi)\lesssim 0.04$ (i.e.\
$\Gamma(\eta\eta')/M \lesssim 0.0005$) is obtained, if one requires that 
prediction for $\Gamma(\pi\pi)/\Gamma(KK)$ remains within the current experimental error bar.\BReep
\end{table}

\begin{table}[p]
\caption[]{Extrapolation of tensor glueball decay\BPR\ with glueball mass
$M=M_T=M_D$ and when the latter is raised to the mass of the tensor glueball candidate $f_2(1950)$, whose experimental width
is 472(18) MeV [$\Gamma/M=0.24(1)$], or
a typical lattice prediction $\sim 2.4 \,{\rm GeV}$, for which the extrapolation from the WSS model predicts an extremely broad tensor glueball.
}
\label{tabtensorextrapolmpi}
\vspace{0.4cm}
\begin{center}
\begin{tabular}{|l|c|c|}
\hline
decay & $M$ & $\Gamma/M$ \\
\hline
$T\to 2\pi$ & 1487 & 0.013\ldots0.018 \\
$T\to 2K$ & 1487 & 0.004\ldots0.006 \\
$T\to 2\eta$ & 1487 & 0.0005\ldots0.0007 \\
$T$ (total) & 1487 & $\approx 0.02\ldots0.03$\\
\hline
$T\to 2\rho\to 4\pi$ & 1944 & 0.129\ldots 0.171 \\ %
$T\to 2K^*\to 2(K\pi)$ & 1944 & 0.105\ldots 0.157 \\
$T\to 2\omega\to 6\pi$ & 1944 & 0.043\ldots 0.057 \\ 
$T\to 2\pi$ & 1944 & 0.014\ldots0.018 \\
$T\to 2K$ & 1944 & 0.009\ldots0.012 \\
$T\to 2\eta$ & 1944 & 0.0017\ldots0.0022 \\
$T$ (total) & 1944 & $\approx 0.30\ldots0.42$\\
\hline
$T\to 2K^*\to 2(K\pi)$ & 2400 & 0.173\ldots 0.250 \\
$T\to 2\rho\to 4\pi$ & 2400 & 0.159\ldots 0.211 \\ %
$T\to 2\omega\to 6\pi$ & 2400 & 0.053\ldots 0.070 \\ 
$T\to 2\phi$ & 2400 & 0.032\ldots 0.051 \\ 
$T\to 2\pi$ & 2400 & 0.014\ldots0.019 \\
$T\to 2K$ & 2400 & 0.012\ldots0.016 \\
$T\to 2\eta$ & 2400 & 0.0025\ldots0.0034 \\
$T\to 2\eta'$ & 2400 & 0.0004\ldots0.0005 \\
$T$ (total) & 2400 & $\approx 0.45\ldots0.62$\\
\hline
\end{tabular}
\end{center}
\end{table}

The Witten-Sakai-Sugimoto model also predicts the decay pattern of tensor glueballs,\cite{Brunner:2015oqa} 
which should in fact be
less sensitive to explicit quark mass terms.
However, the lattice prediction of the tensor glueball mass at around 2.4 GeV is very much above
the holographic result. Extrapolating the latter to such high masses, again by keeping
the dimensionless chiral result of decay into pairs of pseudoscalar mesons fixed,
now gives a very large contribution from decays into $2K^*$, $2\rho$, $2\omega$, and $2\phi$ vector mesons 
as shown in Table \ref{tabtensorextrapolmpi}. This suggests that tensor glueballs in this mass
range may be too broad to be observable at all.
A comparatively broad glueball candidate is provided by the isoscalar $f_2(1950)$ which has $\Gamma/M\approx 0.24(1)$.
With the corresponding mass parameter, the holographic result of 0.3\ldots0.42 appears to be at least marginally comparable.

A more promising case seems to be that of pseudoscalar glueballs. According to the Witten-Sakai-Sugimoto model,
these glueballs should be rather narrow states.\cite{BRinprep}

\section*{Acknowledgments}

The research reported here has been obtained in collaboration with
F.\ Br\"unner and D.\ Parganlija, and has been funded by the Austrian Science Funds through
FWF project no.\ P26366, and the FWF doctoral program no.\ W1252.


\section*{References}

\begin{thebibliography}{99}

\bibitem{Crede:2008vw} 
  V.~Crede and C.~A.~Meyer,
  {\em Prog.\ Part.\ Nucl.\ Phys.\ } {\bf 63}, 74 (2009)
  [arXiv:0812.0600].

\bibitem{Morningstar:1999rf} 
  C.~J.~Morningstar and M.~J.~Peardon,
  {\em Phys.\ Rev.\ }D {\bf 60}, 034509 (1999)
  [hep-lat/9901004].


\bibitem{Lucini:2012gg} 
  B.~Lucini and M.~Panero,
  {\em Phys.\ Rept.\ } {\bf 526}, 93 (2013)
  [arXiv:1210.4997].


\bibitem{Gregory:2012hu} 
  E.~Gregory, {\it et al.}, 
  {\em JHEP} {\bf 1210}, 170 (2012)
  [arXiv:1208.1858].


\bibitem{Witten:1998zw} 
  E.~Witten,
  {\em Adv.\ Theor.\ Math.\ Phys.\  }{\bf 2}, 505 (1998)
  [hep-th/9803131].


\bibitem{Sakai:2004cn} 
  T.~Sakai and S.~Sugimoto,
  {\em Prog.\ Theor.\ Phys.\  }{\bf 113}, 843 (2005)
  [hep-th/0412141].


\bibitem{Sakai:2005yt} 
  T.~Sakai and S.~Sugimoto,
  {\em Prog.\ Theor.\ Phys.\  }{\bf 114}, 1083 (2005)
  [hep-th/0507073].


\bibitem{Rebhan:2014rxa} 
  A.~Rebhan,
  {\em EPJ Web Conf.\ } {\bf 95}, 02005 (2015)
  [arXiv:1410.8858 ].


\bibitem{Brunner:2015oga} 
  F.~Brünner and A.~Rebhan,
  {\em Phys.\ Rev.\ }D {\bf 92}, 121902 (2015)
  [arXiv:1510.07605].


\bibitem{Brower:2000rp} 
  R.~C.~Brower, S.~D.~Mathur and C.-I~Tan,
  {\em Nucl.\ Phys.\ }B {\bf 587}, 249 (2000)
  [hep-th/0003115].


\bibitem{Constable:1999gb} 
  N.~R.~Constable and R.~C.~Myers,
  {\em JHEP} {\bf 9910}, 037 (1999)
  [hep-th/9908175].


\bibitem{Hashimoto:2007ze} 
  K.~Hashimoto, C.-I~Tan and S.~Terashima,
  {\em Phys.\ Rev.\ }D {\bf 77}, 086001 (2008)
  [arXiv:0709.2208].


\bibitem{Brunner:2015oqa} 
  F.~Brünner, D.~Parganlija and A.~Rebhan,
  {\em Phys.\ Rev.\ }D {\bf 91}, 106002 (2015)
  [Erratum: {\em Phys.\ Rev.\ }D {\bf 93}, 109903 (2016)]
  [arXiv:1501.07906].


\bibitem{Narison:1996fm} 
  S.~Narison,
  {\em Nucl.\ Phys.\ }B {\bf 509}, 312 (1998)
  [hep-ph/9612457].


\bibitem{Amsler:1995td} 
  C.~Amsler and F.~E.~Close,
  {\em Phys.\ Rev.\ }D {\bf 53}, 295 (1996)
  [hep-ph/9507326].


\bibitem{Lee:1999kv} 
  W.~J.~Lee and D.~Weingarten,
  {\em Phys.\ Rev.\ D }{\bf 61}, 014015 (2000)
  [hep-lat/9910008].


\bibitem{Janowski:2014ppa} 
  S.~Janowski, F.~Giacosa and D.~H.~Rischke,
  {\em Phys.\ Rev.\ }D {\bf 90}, 114005 (2014)
  [arXiv:1408.4921].


\bibitem{Cheng:2015iaa} 
  H.~Y.~Cheng, C.~K.~Chua and K.~F.~Liu,
  {\em Phys.\ Rev.\ }D {\bf 92}, 094006 (2015)
  [arXiv:1503.06827].


\bibitem{Chanowitz:2005du} 
  M.~Chanowitz,
  {\em Phys.\ Rev.\ Lett.\ } {\bf 95}, 172001 (2005)
  [hep-ph/0506125].


\bibitem{Frere:2015xxa} 
  J.~M.~Frère and J.~Heeck,
  {\em Phys.\ Rev.\ }D {\bf 92}, 114035 (2015)
  [arXiv:1506.04766].


\bibitem{Brunner:2015yha} 
  F.~Brünner and A.~Rebhan,
  {\em Phys.\ Rev.\ Lett.\ } {\bf 115}, 131601 (2015)
  [arXiv:1504.05815].



\bibitem{PDG15}
K.~A. Olive {\it et al.} (Particle Data Group), {\em Chin.\ Phys.\ }C {\bf 38}, 090001 (2014) and 2015 update.

\bibitem{Albaladejo:2008qa} 
  M.~Albaladejo and J.~A.~Oller,
  {\em Phys.\ Rev.\ Lett.\ } {\bf 101}, 252002 (2008)
  [arXiv:0801.4929].


\bibitem{BRinprep}
F.~Brünner and A.~Rebhan, {\em Holographic QCD predictions for production and decay of pseudoscalar glueballs}, in preparation

\end{thebibliography}
\end{document}